# In situ impedance spectroscopy tests of $Li_{4-x}Ge_{1-x}P_xO_4$ as potential solid-state electrolyte for Micro Li-ion Batteries

*Mohammadhossein Montazerian, Kyle J. Stephens, Vladimir Roddatis, Christof Vockenhuber, Arnold Müller, Anders J. Barlow, Thomas Lippert, Nick A. Shepelin[*], Daniele Pergolesi[*]*

M. Montazerian, K. J. Stephens, T. Lippert, N. A. Shepelin, D. Pergolesi

Paul Scherrer Institute PSI, Center for Neutron and Muon Sciences, 5232 Villigen, Switzerland

*daniele.pergolesi@psi.ch

*nikita.shepelin@psi.ch

M. Montazerian, K. J. Stephens, T. Lippert

ETH Zürich, Department of Chemistry and Applied Biosciences, 8093 Zürich, Switzerland

V. Roddatis

GFZ Helmholtz Centre for Geosciences, Telegrafenberg, 14473 Potsdam, Germany

C. Vockenhuber, A. Müller

ETH Zürich, Department of Physics, 8093 Zürich, Switzerland

A. J. Barlow

The University of Melbourne, Materials Characterization and Fabrication Platform (MCFP), 3010 Victoria, Australia

D. Pergolesi

Paul Scherrer Institute PSI, Center for Energy and Environmental Science, 5232 Villigen, Switzerland

*daniele.pergolesi@psi.ch

**Abstract**

Lithium-ion batteries employing solid-state electrolytes (SSEs) are emerging as a safer and more compact alternative to conventional batteries using liquid electrolytes, especially for miniaturized energy storage systems. However, the industry-standard SSE, LiPON, imposes limitations due to its incompatibility with high-temperature processing. In this study, we investigate $Li_{4-x}Ge_{1-x}P_xO_4$ (LGPO), a LISICON-type oxide, as a promising alternative thin-film SSE. LGPO thin films are fabricated using pulsed laser deposition under four distinct deposition conditions, with *in situ* impedance spectroscopy enabling precise conductivity measurements without ambient exposure. We systematically correlate deposition temperature, background pressure, chemical composition, crystallinity, and morphology with ionic transport properties. Polycrystalline LGPO films grown at high temperature (535 °C) and low oxygen pressure (0.01 mbar) exhibited the highest room-temperature ionic conductivity (~$1.2 \times 10^{-5}$ S cm$^{-1}$), exceeding that of LiPON by an order of magnitude, with an activation energy of 0.46 eV. In contrast, amorphous films show significantly lower conductivity (~$5.2 \times 10^{-8}$ S cm$^{-1}$) and higher activation energy (0.72 eV). The results reveal that crystallinity, chemical composition, and grain boundary density critically affect ion transport, highlighting the importance of microstructural control. This work establishes LGPO as a viable, high-performance oxide SSE compatible with high-temperature processing for next-generation microbattery architectures.

## 1. Introduction

Lithium-ion batteries have revolutionized modern energy storage, powering a wide range of applications from consumer electronics to electric vehicles. Their high energy density, long cycle life, and relatively low self-discharge rate have made them the dominant rechargeable battery technology.[1] However, conventional Li-ion batteries rely on liquid electrolytes, which pose safety risks due to flammability and leakage, as well as limitations in operating voltage and stability.[2] To overcome these challenges, solid-state batteries (SSBs) have emerged as a promising alternative, replacing liquid electrolytes with solid-state electrolytes (SSEs). SSEs offer significant advantages over conventional organic liquid electrolytes, including inherent nonflammability and, consequently, enhanced safety.[3,4] Additionally, they exhibit high mechanical strength, which helps suppress the formation of lithium dendrites, offers superior thermal stability, a wider electrochemical stability window, and the potential for formability and miniaturization.[5-7] Despite their considerable promise, SSEs exhibit several critical limitations that impede widespread adoption. A primary obstacle is the high interfacial resistance between solid electrolytes and electrodes, stemming from poor physical contact leading to increased cell impedance and performance degradation over cycles.[8] Additionally, chemical incompatibility can arise, as SSEs often react with lithium metal or high-voltage cathodes, forming resistive interphases that further limit ion transport efficiency.[8] These challenges are compounded by the mechanical brittleness of ceramic electrolytes, which are prone to fracture under cycling-induced stress or volume changes in electrodes, undermining long-term structural integrity.[9]

Thin-film SSEs are a crucial component of thin-film solid-state batteries (TFSSBs), which are considered a promising advancement for next-generation energy storage devices. The reduced dimensions of TFSSBs, coupled with their high power density and rapid charge/discharge rates enabled by the short diffusion pathways for Li ions, make them highly attractive.[10-12] These batteries are particularly valuable for applications requiring high charge/discharge rates, such as drones (during takeoff or landing) and lasers (initial electrical stimulus), or in scenarios where minimal volume is critical, such as wearable sensors, biomedical implants, and active radio-frequency identification (RFID) tags.[13]

Currently, lithium phosphorous oxynitride (LiPON) is widely used as a thin-film SSE due to its wide electrochemical stability window, high electronic resistivity, and sufficiently high ionic conductivity (~$10^{-6}$ S cm$^{-1}$).[14,15] However, its amorphous nature limits its compatibility with high-temperature fabrication processes, as it risks crystallization and consequently losing ionic conductivity during heterostructure formation in TFSSBs. This constraint narrows the range of electrode-electrolyte pairings as high temperature processing is required for many electrode materials. Present TFSSBs utilizing LiPON often rely on lithium metal as the anode, resulting in air sensitivity and complicating the fabrication process. These limitations of LiPON demonstrate the need for alternative thin-film SSEs with sufficiently high ionic conductivity (>$10^{-6}$ S cm$^{-1}$), low electronic conductivity, stability during high-temperature processing (>400 °C), and the ability to ensure long cycle life (1000 cycles) and reliable performance in TFSSBs.[16]

Among the various inorganic materials considered for SSEs—such as sulfides, oxides, and halides—oxide-based electrolytes stand out due to their superior thermal stability, broader electrochemical stability windows suitable for high-voltage cathodes, and lower sensitivity to moisture.[17, 18] However, oxides generally exhibit lower ionic conductivity compared to their sulfide and halide counterparts.[17]

Oxide-based SSEs can be categorized into several structural families, including perovskite, garnet, lithium super ionic conductor (LISICON), and sodium super ionic conductor (NASICON) types.[17, 18] Each of these material groups has demonstrated potential for use in SSEs. Examples of promising oxide SSEs are the perovskite-based $Li_{3x}La_{2/3-x}TiO_3$ (LLTO), the garnet-based $Li_7La_3Zr_2O_{12}$ (LLZO), the LISICON-based $Li_{4-x}Ge_{1-x}P_xO_4$ (LGPO), and the NASICON-based $Li_{1+x}Al_xTi_{2-x}(PO_4)_3$ (LATP). Thin films of LLTO have been deposited using pulsed laser deposition (PLD), achieving an ionic conductivity of ~$10^{-6}$ S cm$^{-1}$ at room temperature. [14] Similarly, LLZO thin films prepared via sputtering have demonstrated an ionic conductivity of ~$10^{-4}$ S cm$^{-1}$, while LGPO and LATP films grown using PLD exhibit an ionic conductivity of ~ $10^{-6}$ S cm$^{-1}$ and ~5 × $10^{-7}$ S cm$^{-1}$ at room temperature, respectively.[15,17,19] However, the deposition temperatures required for LLTO, LLZO, and LATP exceed 700–850 °C, which can induce interdiffusion of elements in neighboring layers of the microbattery structure.[20-22] Additionally, the deposition of these materials is challenging due to the tendency for insulating secondary phases to form, such as $La_2Ti_2O_7$ and $La_2Zr_2O_7$, respectively, further complicating their processing.[20-23] **Table 1** summarizes this information about different oxide-based SSEs.

**Table 1.** Overview of room-temperature conductivity, deposition techniques and challenges for common oxide-based solid-state electrolytes[14,15,17,19]

| SSE | Thin film room-temperature conductivity (S cm$^{-1}$) | Thin film deposition technique | Thin film deposition challenges |
|---|---|---|---|
| LLTO | ~10$^{-6}$ | PLD | High deposition temperature, La$_2$Ti$_2$O$_7$ secondary phase, air sensitive |
| LLZO | ~10$^{-4}$ | Radio frequency sputtering | High deposition temperature, La$_2$Zr$_2$O$_7$ secondary phase, air sensitive |
| LGPO | ~10$^{-6}$ | PLD | No challenges, wide deposition temperature, non-air sensitive |
| LATP | ~5 × 10$^{-7}$ | PLD | High deposition temperature |
| LiPON | ~10$^{-6}$ | Radio frequency sputtering | Can not be deposited above 400 °C |

The LISICON family is particularly notable for its structural flexibility and the tunability of ionic conductivity and activation energy through cation and anion substitution.[24] The LISICON structure is derived from γ-Li$_3$PO$_4$, with lithium and phosphorus occupying half of the tetrahedral sites within a distorted hexagonal close-packed framework. Subvalent cation substitution of phosphorus in Li$_3$PO$_4$ introduces charge imbalances that are compensated by additional lithium ions occupying interstitial octahedral sites. Alternatively, supervalent substitution of metal cations (M) in oxides such as Li$_4$MO$_4$ (M = Si, Ge, Ti) results in LISICON-type materials of the form Li$_{4-y}$M$_{1-x}$M'$_x$O$_4$. In these structures, three lithium ions contribute to the tetrahedral framework, while the remaining ions (1-y) occupy interstitial sites, enhancing ionic mobility. Partial substitution of Ge$^{4+}$ with P$^{5+}$ results in a solid solution of γ-Li$_3$PO$_4$ and Li$_4$GeO$_4$ with the formula Li$_{4-x}$Ge$_{1-x}$P$_x$O$_4$ (LGPO). Previous studies on LGPO pellets have investigated the effect of LGPO's chemical composition on its ionic conductivity, demonstrating that the aliovalent substitution of P$^{5+}$ with Ge$^{4+}$ enhances ionic conductivity by 4–5 orders of magnitude compared to its parent compounds.[24,25] Furthermore, the activation energy decreases to approximately 0.5 eV, which has been attributed to a reduction in the formation energy of mobile charge carriers.[24]

Additionally, it has been reported that LGPO can be successfully deposited as thin films using PLD, with its thin-film conductivity aligning well with that of its bulk counterpart.[19] However, due to the limited crystallinity in deposited films, the impact of crystallinity on LGPO's ionic conductivity has not been thoroughly explored. This relationship is also difficult to investigate using pellets, as they do not exhibit well-defined crystalline orientations. Understanding this correlation is particularly important for technological applications, especially in micro Li ion batteries, where optimizing the crystallinity-conductivity relationship is crucial. Therefore, this study provides valuable insights into the role of crystallinity in LGPO thin films, bridging an important gap in the field.

To understand the Li ion behavior in this member of the LISICON family suitable for technological applications, this study investigates the correlation between deposition conditions, crystallinity and crystallographic properties, and the ion mobility characteristics in the LGPO thin films. In order to achieve this, we have grown LGPO thin films using PLD, varying the deposition temperature and background gas pressure. Paying particular attention to possible degradation commonly observed in Li-containing materials, the Li ion conductivity of the films was probed using *in situ* impedance spectroscopy, which enabled analysis immediately after growth in a controlled atmosphere without air exposure. We observe that crystalline and morphological features of the film significantly affect the ionic conductivity. The observed conductivity values are the largest to date in a thin film system, and are in line with bulk samples, exceeding the room temperature conductivity of LiPON by one order of magnitude. Thus, such values align well with existing solid state Li ion conducting materials, while allowing the possibility of high temperature growth, which facilitates the realization of oxide TFSSBs.

## 2. Results and discussion
### 2.1. LGPO thin films structural, morphological, and chemical characterization

Four sets of samples were used for this study: high temperature and high pressure (HTHP), High temperature and low pressure (HTLP), intermediate temperature and low pressure (ITLP), and low temperature and low pressure (LTLP). **Table 2** reports the deposition parameters. To evaluate the ionic transport behavior of the films, *in situ* impedance spectroscopy, its schematic can be observed in **Figure 1**, was conducted for HTLP and LTLP samples across a temperature range of 535–350 °C and 150–400 °C, respectively.

Table 2. Deposition conditions for the prepared LGPO samples

|  | Deposition temperature (°C) | Deposition pressure (m bar) |
|---|---|---|
| High T, Low P (HTLP) | 535 | 0.01 |
| Intermediate T, Low P (ITLP) | 350 | 0.01 |
| Low T, low P (LTLP) | 25 | 0.01 |
| High T, High P (HTHP) | 535 | 0.05 |

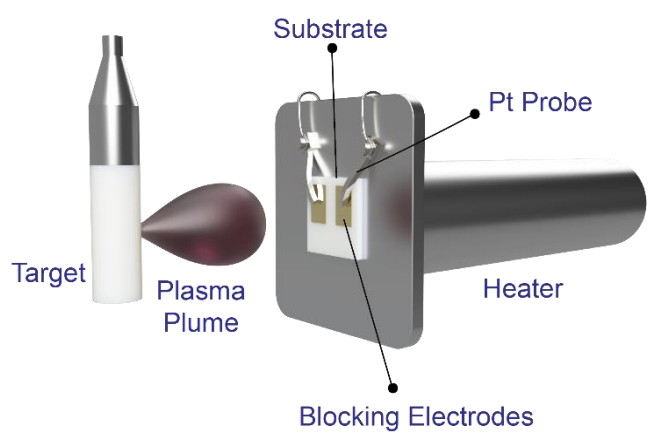

Figure 1. Schematic image of the *in situ* impedance spectroscopy set-up.

The XRD patterns of the samples are shown in **Figure 2a**. The HTHP and ITLP samples exhibit a single out-of-plane crystallographic orientation within the measured range, corresponding to the (002) and (410) planes, respectively. In contrast, the HTLP sample displays multiple out-of-plane reflections, confirming its polycrystalline nature. The as-grown LTLP sample shows no distinct diffraction peaks, indicating that it is predominantly amorphous, with potential nanoscale crystallites dispersed throughout the film. Interestingly, the LTLP sample, after being heated to 400 °C during impedance measurements, reveals two weak diffraction peaks, suggesting a structural transition from an amorphous to a partially polycrystalline phase.

In order to gain more insights about the local crystal structure of the films, selected area electron diffraction (SAED) patterns of these samples have been collected and are shown in **Figure 2b-e**. **Figure 2b**, corresponding to sample HTHP, exhibits a pattern with sharp and well-defined spots corresponding to a single crystallographic orientation. The d-spacing is ~5.0

Å which is in a good agreement with (002) crystal planes in *P-n-m-a* space group (5.02 Å according to ICSD_250066).

**Figure 2c** shows rings composed of discrete spots indicating the presence of multiple orientated grains confirming the polycrystalline nature of the film.

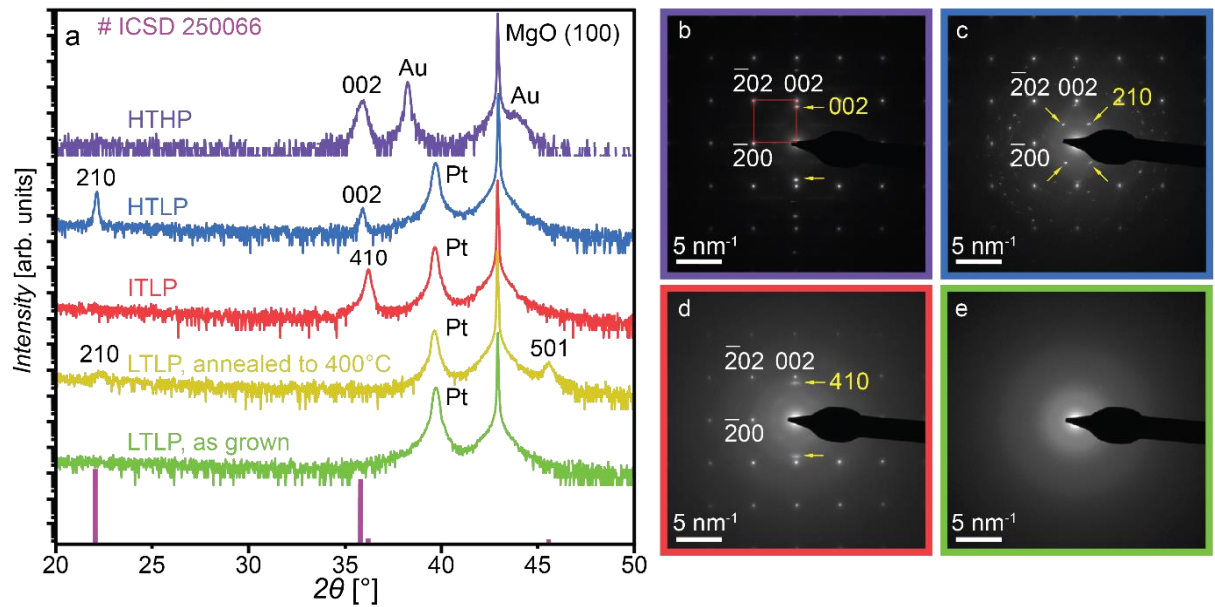

**Figure 2.** (a) X-ray diffractograms of the LGPO films, Electron diffraction patterns of LGPO films (b) HTHP, (c) HTLP, (d) ITLP, (e) LTLP. Sharp indexed spots forming squared pattern originate from MgO substrate. The spots pointed with yellow arrows originate from the LGPO films. For the LTLP (e) only diffraction from the film is shown.

The SAED pattern in **Figure 2d** displays multiple diffraction rings, confirming a polycrystalline nature of the deposited thin film. The presence of distinct diffraction spots along the rings with well-defined symmetry suggests that the crystallites are relatively well-defined out-of-plane, albeit randomly oriented in-plane. The measured d-spacing is 2.49 Å which is in agreement with that of the (410) crystal plane (2.45 Å according to ICSD_250066). Therefore, this sample is characterized as a textured polycrystalline film. In contrast, **Figure 2e** presents a diffuse halo typical of an amorphous material.

SEM and HIM micrographs of the three crystalline samples (shown in **Figure 3**) were analyzed using ImageJ software to determine their average grain sizes, with the results summarized in **Table 3**. The analysis indicates that the HTLP and ITLP samples possess similar average grain sizes, whereas the HTHP samples exhibit smaller grains, approximately half the size of those in the HTLP and ITLP samples.

The chemical composition of the four samples, determined by heavy ion elastic recoil detection analysis (HI-ERDA)[26], is also presented in **Table 3**. The compositional analysis provides insights into the lithium content and the degree of $Ge^{4+}$ substitution by $P^{5+}$, i.e., the value of x. **Table 3** reveals that the HTLP and HTHP samples exhibit very similar chemical compositions. Additionally, while the ITLP and LTLP samples show the same level of $Ge^{4+}$ substitution by $P^{5+}$, they differ in their lithium content. It is noteworthy that when comparing the LTLP, ITLP, and HTLP samples, an increase in deposition temperature correlates with an increase in lithium content. This trend is consistent with previous reports on LGPO thin films.[19]

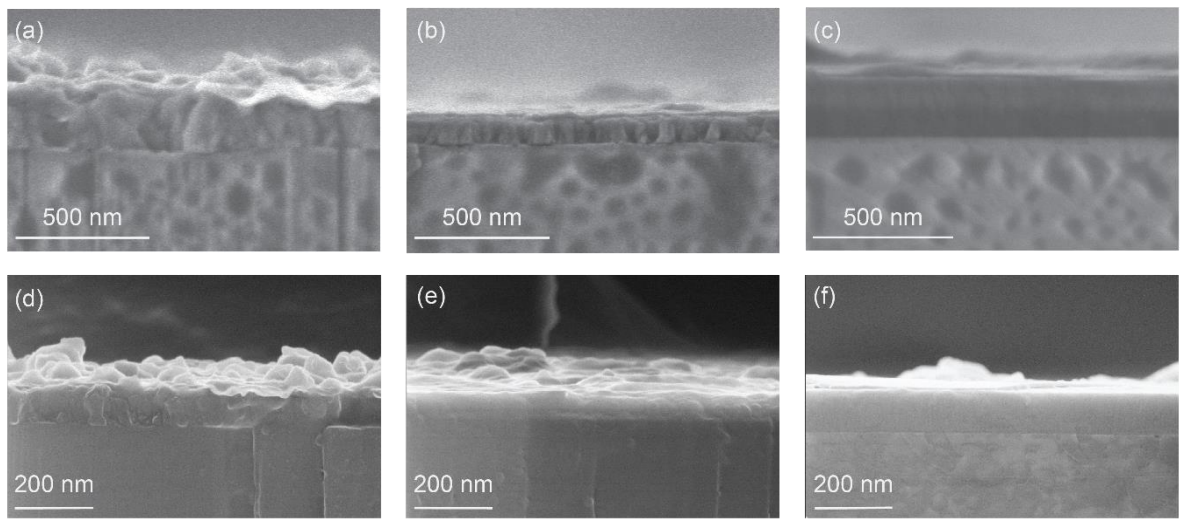

**Figure 3.** The SEM micrograph of (a) HTLP, (b) HTHP, (c) ITLP samples, HIM micrographs of (d) HTLP, (e) HTHP, (f) ITLP samples.

## 2.2. LGPO thin films electrical characterization

It is well known that Li content, degree of crystallinity, and average grain size play a crucial role in determining the ionic transport properties of a material. These factors will be further correlated with impedance spectroscopy measurements to assess their impact on ionic conductivity. The measured ionic conductivities of the samples at 400 °C, along with their chemical composition and average grain size, are presented in **Table 3**.

**Table 3.** LGPO samples' chemical compositions, average grain size, and ionic conductivities at 400 °C

| Sample | Chemical composition | Average grain size (nm) | σ (S cm$^{-1}$) at 400 °C |
|---|---|---|---|
| HTLP | Li$_{3.08}$ Ge$_{0.52}$ P$_{0.47}$ O$_4$ | 193 ± 15.3 | 0.24 ± 0.0018 |
| ITLP | Li$_{2.96}$ Ge$_{0.72}$ P$_{0.32}$ O$_4$ | 179 ± 32.3 | 0.0056 ± 0.0024 |
| LTLP | Li$_{2.44}$ Ge$_{0.72}$ P$_{0.41}$ O$_4$ | Amorphous | 0.021 ± 0.0012 |
| HTHP | Li$_{3.08}$ Ge$_{0.56}$ P$_{0.48}$ O$_4$ | 84.0 ± 13.0 | Open circuit [1)] |

[1)] No measured conductivity with the available electronics;

Among the tested samples, the highest lithium content is observed in those deposited at high temperatures (HTLP and HTHP), whereas the lowest lithium content is found in the samples deposited at low temperature (LTLP). However, the ionic conductivity trend does not directly correlate with lithium content, suggesting that Li concentration alone cannot account for the observed variations in conductivity. Therefore, other structural or morphological factors must be considered.

From the SAED patterns (**Figure 2**), it is evident that while the HTLP and ITLP samples are both polycrystalline, the ITLP samples exhibit a textured out-of-plane orientation. Despite their similar grain sizes, the ITLP samples exhibit an ionic conductivity that is 40 times lower than that of the HTLP samples. This strong discrepancy suggests that crystallinity and texture significantly influence in-plane ionic transport. This assumption is further supported by the HTHP samples for which the SAED pattern shows a well-defined crystal orientation out-of-plane. Although the HTHP samples show a very similar Li content compared to HTLP and ITLP, the ionic conductivity of the HTHP samples is orders of magnitude lower and below the measurement capability of the Biologic VMP 300 potentiostat.

In addition, the impact of the average grain size on the ionic conductivity can be observed by comparing the two out-of-plane textured films, HTHP and ITLP. The average grain size of the HTHP samples is almost half of the ITHP samples which leads to higher grain boundary contribution to the total electrical resistance in the HTHP samples. Since the total ionic conductivity of ITLP samples is orders of magnitude higher than that of HTHP samples, we conclude that the grain boundary regions can dramatically affect the electrical properties of LGPO.

Now that the influence of crystallinity and morphology on the ionic conductivity of LGPO thin films has been established, we turn to evaluating the activation energy for ion conduction in both polycrystalline and amorphous LGPO films. For the HTLP samples, temperature-dependent resistance measurements were carried out between 535 °C and 350 °C, beginning at the deposition temperature (535 °C) and decreasing stepwise. A similar measurement sequence was performed for the LTLP samples over the range of 150 °C to 400 °C, starting from room temperature and increasing stepwise. At each temperature step, the sample was equilibrated for 30 minutes prior to measurement. For the LTLP and ITLP samples, resistance measurements were obtained only at approximately 400 °C. At this temperature, the HTHP samples exhibited electrical resistance values that were too high to be reliably measured.

**Figure 4** presents the Nyquist plots of the HTLP and LTLP films measured at various temperatures. The equivalent circuit model used to fit the HTLP data (inset of **Figure 4 (a)**) includes: (1) a series resistor ($R_{external\ circuit}$) representing the resistance of the external circuit and the contact resistance at the ion-blocking electrodes; (2) a parallel combination of a resistor and a constant phase element ($Q_{Grain\ +\ GB}/R_{Grain\ +\ GB}$) corresponding to the combined contributions of the grain interiors and grain boundaries within the LGPO film; and (3) a constant phase element ($Q_{electrodes}$) accounting for charge polarization at the electrode–electrolyte interfaces.

For the LTLP samples, the equivalent circuit (inset of **Figure 4 (b)**) excludes the grain boundary contribution, as the amorphous structure of the film lacks distinct grains. The only exception is the measurement at 400 °C, where, as previously discussed, the film undergoes crystallization, and the grain boundary contribution becomes present.

Due to the low thickness of the films (100 to 250 nm), the stray capacitance from the MgO substrate (~500 µm thick) dominates, preventing a distinction between bulk and grain boundary contributions.[27]

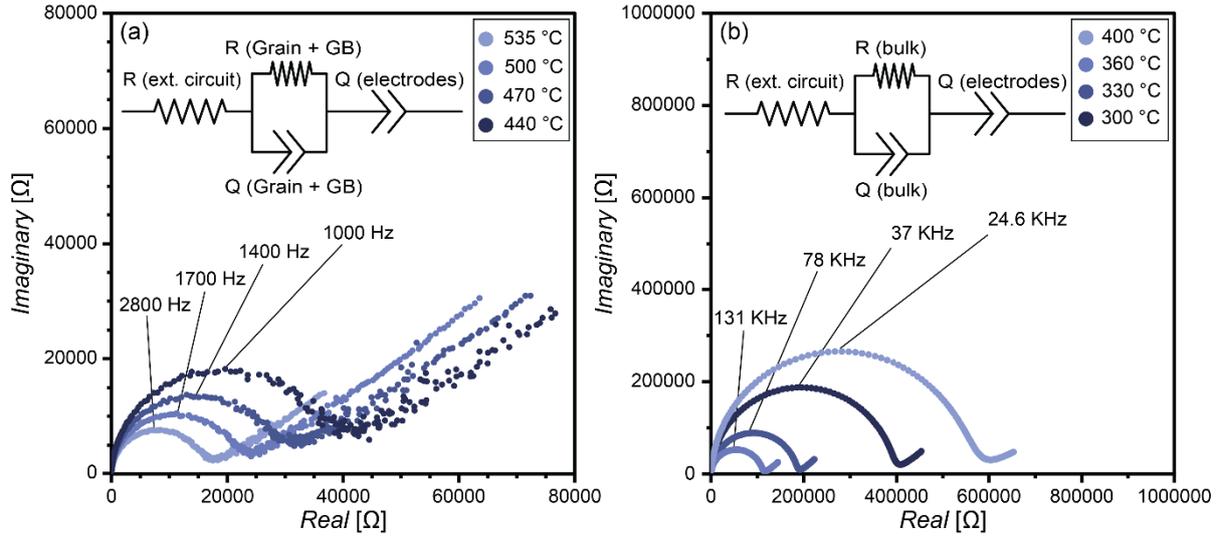

**Figure 4.** (a) The Nyquist plots of the LGPO sample grown at high temperature and low pressure (HTLP), and (b) low temperature and low pressure (LTLP).

The Arrhenius plot for the LGPO thin films is depicted in **Figure 5**. The ionic conductivity measurement for the ITLP and HTHP samples is only shown at 400 °C as this was the lowest temperature possible to reliably measure the conductivities of these samples with the electronics available for these measurements. The extracted activation energy ($E_a$) of the HTLP sample is 0.46 eV, and its extrapolated room-temperature ionic conductivity ($\sigma_{RT}$) is ~$1.2 \times 10^{-5}$ S cm$^{-1}$.

The typical room-temperature ionic conductivity reported for LiPON, the benchmark SSE for thin-film batteries, ranges from $7.46 \times 10^{-7}$ S·cm$^{-1}$ to $4.9 \times 10^{-6}$ S·cm$^{-1}$ (the black bar shown in **Figure 5**).[28] In comparison, our polycrystalline LGPO sample exhibits approximately an order of magnitude higher conductivity, highlighting its potential as a promising alternative to LiPON for use in TFSSBs.

Moreover, looking at the plot of the LTLP samples, we see a continuous line from 150 °C to 360 °C, but towards 400 °C, we see a drop in the conductivity of the samples. This is the sign that in this temperature range (360 °C to 400 °C), crystallization of the films initiates and the ion conduction mechanism alters. When the temperature is reduced back to 150 °C after crystallization, the conductivity remains lower than that of the original amorphous phase. The activation energy ($E_a$) extracted for the LTLP sample in the 150 °C to 360 °C range is 0.72 eV, with an extrapolated room-temperature ionic conductivity of approximately $5.2 \times 10^{-8}$ S cm$^{-1}$.

The chemical composition of the HTLP films shown in **Table 3**, is $Li_{3.08}Ge_{0.52}P_{0.47}O_4$, which corresponds to x ≈ 0.5 (in $Li_{4-x}Ge_{1-x}P_xO_4$). The $E_a$ and $\sigma_{RT}$ values of HTLP samples are consistent with previous studies on LGPO pellets. For instance, Rodger et al.[25] have reported a room-temperature conductivity of ~$10^{-5}$ S cm$^{-1}$ for LGPO pellets with x = 0.5, while Song et al.[29] have obtained $3.8 \times 10^{-6}$ S cm$^{-1}$ with $E_a$ = 0.5 eV for x = 0.66.

Prior studies on LGPO pellets with Ge content between 0.4 and 0.6 have reported a maximum $\sigma_{RT}$ = $1 \times 10^{-5}$ S cm$^{-1}$ and $E_a$ = 0.5 – 0.6 eV.[24] These findings support the hypothesis that a higher $Ge^{4+}$ content ((1-x) relative to that of $P^{5+}$ (x)) increases the unit cell volume, facilitating Li-ion migration via larger interstitial sites. However, at higher $Ge^{4+}$ contents (1-x > 0.6), an increase in activation energy occurs due to the inductive effect, wherein $Ge^{4+}$ (less positively charged than $P^{5+}$) reduces electron withdrawal from oxygen, resulting in higher effective charges on oxygen and therefore stronger electrostatic interactions with $Li^+$ ions.[24] Consequently, while the pre-exponential factor ($\sigma_o$) in Equation 1 increases with Ge content, the net RT conductivity remains relatively unchanged. This can in fact explain the higher activation energy of the LTLP samples as its chemical composition is $Li_{2.44}Ge_{0.72}P_{0.41}O_4$ meaning that its Ge content is beyond 0.6 (x ≈ 0.3).

$$\sigma_T = \sigma_0 \exp\left(\frac{-E_a}{KT}\right) \qquad (1)$$

Furthermore, Gilardi et al.[19] have investigated LGPO pellets and thin films, reporting $E_a$ = 0.53 eV and $\sigma_{RT}$ = ~$10^{-6}$ S cm$^{-1}$ for LGPO pellet with x = 0.8. The polycrystalline thin films in this study (grown at 500 °C) with chemical composition of $Li_{2.7\pm0.3}Ge_{0.49}P_{0.5}O_{3.7\pm0.3}$ (x ≈ 0.5) exhibited $\sigma_{RT}$ = ~$3 \times 10^{-6}$ S cm$^{-1}$ and $E_a$ = 0.51 eV. The activation energy value is very in line with the HTLP samples (0.46 eV) which has chemical composition of $Li_{3.08}Ge_{0.52}P_{0.47}O_4$ (x ≈ 0.5). Although, the room-temperature conductivity of HTLP samples is an order of magnitude higher due to higher Li content and better crystal structures. Moreover, Gilardi et al.[19] reported amorphous thin films with and $E_a$ = 0.58 eV and $\sigma_{RT}$ = ~$3 \times 10^{-7}$ S cm$^{-1}$ which have chemical composition of $Li_{1.58\pm0.4}Ge_{0.45}P_{0.55}O_{3.27\pm0.5}$ (x ≈ 0.5). Our LTLP samples with chemical composition of $Li_{2.44}Ge_{0.72}P_{0.41}O_4$ (x ≈ 0.3) shows $E_a$ = 0.72 eV and $\sigma_{RT}$ = ~$5.2 \times 10^{-8}$ S cm$^{-1}$. The discrepancy between LTLP samples and Gilardi et al.[19] amorphous samples activation energy and room-temperature conductivity is due to the difference in their chemical compositions.

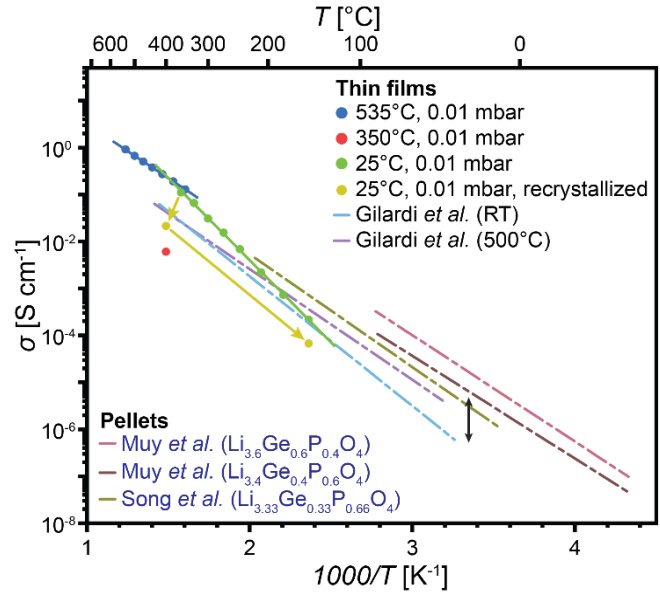

**Figure 5.** The Arrhenius plot of HTLP sample obtained with *in situ* impedance spectroscopy set-up, LGPO thin films deposited at 500 °C and RT measured by Gilardi *et al.*[19] and LGPO pellets measured by Muy *et al.*[24] and Song *et al.*[29]. The black vertical bar shows the room-temperature conductivity range reported for LiPON.[28]

## 3. Conclusion

In this study, we have systematically investigated the ionic conductivity of LGPO thin films grown via PLD under varying deposition conditions. The primary objective was to assess the influence of deposition parameters on the crystallinity, morphology and ionic transport properties of the films. *In situ* impedance spectroscopy was employed for precise conductivity characterization, minimizing external contamination effects.

Our findings indicate that the ionic conductivity and activation energy values obtained using the *in situ* setup are in strong agreement with previous studies on LGPO. However, the higher precision of our measurements—achieved by preventing air exposure and thereby avoiding carbonate formation—may account for the small discrepancies observed when compared to earlier reports on LGPO thin films.[19] This highlights the importance of environmental control in accurately evaluating the electrochemical properties of thin-film solid electrolytes.

Furthermore, the impact of Li content, crystallinity, and grain size on the ionic conductivity of LGPO films was analyzed. Interestingly, while the high temperature (535 °C) and low pressure (0.01 mbar) and intermediate temperature (350 °C) and low pressure (0.01 mbar) samples exhibited similar Li content and grain sizes, their ionic conductivities differed by a factor of 40. This significant variation suggests that a textured crystalline orientation leads to increased grain boundary resistance, thereby reducing in-plane ionic transport.

This hypothesis is further reinforced by the behavior of the high temperature (535 °C) and high pressure (0.05 mbar) sample, which also exhibits textured crystallinity and an even smaller grain size, yet has a similar Li content. Despite this, it demonstrated no measurable in-plane conductivity at 400 °C. These findings confirm that both crystallographic texture and grain boundary density play a critical role in determining the ionic transport properties of LGPO thin films.

The ionic conductivity values obtained for polycrystalline LGPO films demonstrate that LGPO is a versatile solid-state electrolyte. Its high room-temperature ionic conductivity (~$1.2 \times 10^{-5}$ S cm$^{-1}$) highlights its potential as a compelling alternative to the widely used LiPON (~$10^{-6}$ S cm$^{-1}$) in micro-scale solid-state batteries, enabling high-temperature fabrication processes and broadening the design space for next-generation microbattery architectures. On the other hand, amorphous LGPO with high activation energy (0.72 eV) and very low room-temperature conductivity (~$5.2 \times 10^{-8}$ S cm$^{-1}$) is not an alternative for thin film batteries. However, optimization of the chemical composition has been shown to improve the room-temperature ionic conductivity of amorphous LGPO, suggesting potential for future enhancement.[19]

## 4. Experimental Methods
### 4.1. LGPO Powder and PLD Target Fabrication

To synthesize LGPO powder via a solid-state reaction, stoichiometric amounts of lithium carbonate (Li$_2$CO$_3$) and germanium oxide (GeO$_2$) were mixed according to reaction (1):

$$2\,Li_2CO_3 + GeO_2 \rightarrow Li_4GeO_4 + 2CO_2 \qquad (1)$$

The powders were thoroughly ground using an agate mortar and pestle for 30 minutes. The resulting mixture was heated in a tubular furnace at 800 °C for 8 hours in air, with a heating and cooling rate of 5 °C/min. Following this, stoichiometric amounts of Li$_3$PO$_4$ were added to the synthesized Li$_4$GeO$_4$ powder based on reaction (2):

$$Li_4GeO_4 + 4Li_3PO_4 \rightarrow 5Li_{3.2}P_{0.8}Ge_{0.2}O_4 \qquad (2)$$

This powder mixture was then subjected to heating at 900 °C for 12 hours under a constant oxygen flow in a tubular furnace, with a ramping rate of 5 °C/min.

A common challenge when growing Li-containing films *via* pulsed laser deposition is the interaction of the ablated species with the gaseous environment during their time of flight, which leads to preferential compositional deficiencies of the lighter elements. [30] To account for lithium loss during PLD, 5 mol% excess $Li_2O$ was added to the synthesized LGPO powder. The mixture was pressed into a dense pellet using a uniaxial hydraulic press at 5 bar. The resulting pellet was sintered at 900 °C in air for 12h with heating/cooling rate of 5 °C.min$^{-1}$.

## 4.2. LGPO Thin Film Growth

LGPO thin films were deposited using PLD. Prior to film deposition, two rectangular blocking electrodes for Li-ions (Au or Pt, 100 nm thick) were sputtered/e-beam evaporated onto MgO (100) substrates through a shadow mask, maintaining a 1 mm gap between electrodes. The substrate was glued onto the substrate holder using silver paint and mounted in a vacuum chamber.

A 248 nm KrF excimer laser operating at 10 Hz, spot size of 0.0108 cm$^2$, fluence of 2.1 J cm$^{-2}$, and oxygen as the background gas was used for deposition, with the target-substrate distance set to 6 cm. During deposition, Pt probes connected to the read-out electronics were placed in contact with the blocking electrodes to enable *in situ* impedance spectroscopy measurements.

Four deposition conditions were investigated for this study, where we modulated the substrate temperature (T) and background gas pressure (P). The grown samples were prepared at: (1) a high temperature (535 °C) and low pressure (0.01 mbar); (2) a high temperature (535 °C) and a high pressure (0.05 mbar); (3) an intermediate temperature (350 °C) at low pressure (0.01 mbar); and (4) low temperature (25 °C) at low pressure (0.01 mbar). A summary of the deposition conditions is shown in **Table 2**.

## 4.3. Thin Film Characterization

### 4.3.1. Crystal Structure Characterization

*X-Ray diffraction*

The crystal structure of the LGPO films was analyzed using X-ray diffraction (XRD). Out-of-plane ω-2θ scans were performed on a Bruker D8 Discover instrument with monochromatic Cu K$\alpha_1$ radiation. Data was collected over a 2θ range of 20°–50° with a step size of 0.02°.

*Electron diffraction*

Cross-section specimens for Selected Area Electron Diffraction (SAED) studies were prepared by a lift-out technique using a Thermo Fischer Scientific (TFS) Helios G4 UC DualBeam (FIB-SEM) instrument. SAED patterns were collected using a TFS Themis Z (3.1) Scanning TEM (STEM) operated at 300kV. The microscope is equipped with a TFS SuperX$^{TM}$ Energy dispersive X-ray (EDX) spectrometer as well as with a Gatan Continuum 1065ER Electron Energy Loss Spectrometer (EELS). The surface of MgO substrate has roughness of ~5 nm. The LGPO films is easily damaged and decomposed by electron (and ion) beam.

### 4.3.2. Morphological Characterization

The morphology of the LGPO films were examined using scanning electron microscopy (SEM, Hitachi Regulus 8230) and He-ion beam microscopy (HIM, Zeiss ORION NanoFab). Imaging with SEM was conducted in secondary electron mode under vacuum using an acceleration voltage of 2 kV, with a working distance of approximately 3.5 mm. HIM imaging was performed with a 30 keV He$^+$ beam at nominally 0.4 pA beam current at approximately 9.4 mm working distance. Secondary electron images were collected and *in situ* surface charge neutralisation was performed via low-energy electron flooding during acquisition.

### 4.3.3. Compositional Characterization

Elemental depth profiling was performed using heavy ion elastic recoil analysis (HI-ERDA) at the Laboratory of Ion Beam Physics, ETH Zürich. A time-of-flight and energy (ToF-E) spectrometer measures the velocity and energy of the recoil ions produced by elastic collisions with 13 MeV beam under a total scattering angle of 36°.[26]

## 4.4. In situ Impedance Spectroscopy

Impedance spectroscopy measurements were conducted using a Biologic VMP 300 potentiostat and an Agilent E4980a LCR meter. An AC potential was applied via 0.25 mm-thick Pt needle probes in contact with the Au or Pt blocking electrodes. The in-plane conductivity (i.e., parallel to the substrate surface) of the films was measured under a controlled oxygen atmosphere (0.01 mbar) within the PLD chamber. Impedance data was collected over the frequency range between 1 Hz to 2 MHz with a voltage amplitude of 50 mV to 500 mV. Data was analyzed using EC-Lab software to extract the resistance of the films. A schematic of the *in situ* impedance spectroscopy setup is shown in **Figure 1**.

## Acknowledgment

The authors acknowledge financial support from the Swiss National Science Foundation (SNSF) through the project scheme (204103) and Ambizione scheme (216196). This work was performed in part at the Materials Characterisation and Fabrication Platform (MCFP) at the University of Melbourne and the Victorian Node of the Australian National Fabrication Facility (ANFF). In addition, authors would like to thank the European Regional Development Fund and the State of Brandenburg for the Themis Z (3.1) microscope (part of Potsdam Imaging and Spectral Analysis (PISA) facility).

## Conflict of Interest

All authors approve that there is no conflict of interest.

## Data availability

All data supporting the conclusions in the paper are present in the paper. Raw data and metadata are available in the following repository: *"URL: https://doi.org/10.5281/zenodo.15807730"*.